\documentclass[a4paper,onecolumn,accepted=2025-03-31]{quantumarticle}
\pdfoutput=1
\usepackage[letterpaper,hmargin=1in,vmargin=1in]{geometry} 
\usepackage{fullpage}
\usepackage[pdfstartview=FitH,colorlinks,linkcolor=blue,citecolor=blue]{hyperref} 

\usepackage[T1]{fontenc}
\usepackage{amsthm} 
\usepackage{color} 

\usepackage{microtype,pdfsync} 
\usepackage{amsmath,amsfonts,amssymb}
\usepackage{url} 
\usepackage{xspace,paralist}

\usepackage{float}
\floatstyle{boxed}
\restylefloat{figure}
\usepackage{multirow}

\usepackage{xfrac}

\theoremstyle{plain} 

\newtheorem{theorem}{Theorem}[section] 
\newtheorem{lemma}[theorem]{Lemma} 
 
\newtheorem{corollary}[theorem]{Corollary}

\theoremstyle{definition} 

\newtheorem{definition}[theorem]{Definition}

\newcommand{\Ident}{{\mathbf{I}}}

\newcommand{\Om}{{\mathbf{O}}}

\newcommand{\Um}{{\mathbf{U}}}

\newcommand{\As}{{\mathcal{A}}}
\newcommand{\Bs}{{\mathcal{B}}}

\newcommand{\Hs}{{\mathcal{H}}}

\newcommand{\Qs}{{\mathcal{Q}}}

\newcommand{\Xs}{{\mathcal{X}}}
\newcommand{\Ys}{{\mathcal{Y}}}

\newcommand{\adv}{{\As}}
\newcommand{\advB}{{\Bs}}

\newcommand{\OWF}{{\sf OWF}}
\newcommand{\PRF}{{\sf PRF}}
\newcommand{\PRP}{{\sf PRP}}
\newcommand{\FPC}{{\sf F}}
\newcommand{\FPCInv}{{\sf R}}

\newcommand{\Comp}{{\sf Comp}}
\newcommand{\CComp}{{\sf ClassicComp}}
\newcommand{\QComp}{{\sf QuantComp}}
\newcommand{\Query}{{\sf Query}}
\newcommand{\CQuery}{{\sf ClassicQuery}}
\newcommand{\QQuery}{{\sf QuantQuery}}
\newcommand{\Domain}{{\sf Dom}}
\newcommand{\LargeD}{{\sf LargeDom}}
\newcommand{\AnyD}{{\sf AnyDom}}

\newcommand{\poly}{{\sf poly}}
\newcommand{\negl}{{\sf negl}}

\newcommand{\qstate}[2]{{\left|\psi_{#1}^{#2}\right\rangle}}

\newcommand{\ignore}[1]{}

\floatstyle{plain}
\restylefloat{figure}

\begin{document}

\title{A Note on Quantum-Secure PRPs}
\author{{\sc Mark Zhandry}}
\email{mzhandry@gmail.com}
\thanks{This paper was initially written while at Princeton University, but the idea had been in my head since I was a PhD student at Stanford University.}
\affiliation{NTT Research}

\date{}
\maketitle

\begin{abstract} We show how to construct pseudorandom permutations (PRPs) that remain secure even if the adversary can query the permutation, both in the forward and reverse directions, on a quantum superposition of inputs.  Such quantum-secure PRPs have found numerous applications in cryptography and complexity theory. Our construction combines a quantum-secure pseudorandom \emph{function} together with constructions of \emph{classical} format preserving encryption.  By combining known results, we show how to construct quantum-secure PRP in this model whose security relies only on the existence of one-way functions. 
\end{abstract}

\section{Introduction}

\label{sec:intro}

Pseudorandom permutations (PRPs), a formalization of block ciphers, are one of the most widely used cryptographic building blocks, as they underlie most symmetric key encryption used today. A PRP is a classical, efficiently computable keyed permutation that looks like a truly random permutation to anyone that is only given oracle access to the function.  Additionally, given the key it is also possible to efficiently invert the permutation. A celebrated foundational result is that PRPs can be built from a much weaker tool --- a one-way function, which is a classical function that is easy to compute but hard to invert.  

\paragraph{Enter quantum computers.}  In this work, we consider a very strong \emph{quantum} adversary model for PRPs.  Namely, we allow the adversary to not only poses a quantum computer, but also make quantum superposition queries to the permutation, both in the forward and reverse direction.  Such PRPs are a very natural object in the quantum setting, and have found numerous applications (see Section~\ref{sec:apps} below for examples).

We investigate the following open question raised by Zhandry~\cite{FOCS:Zhandry12}: whether or not such strong PRPs can still be built from (quantum resistant) one-way functions as in the classical case.  We resolve this question positively:
\begin{theorem}\label{thm:main}Assuming quantum resistant one-way functions exist, so do quantum-secure PRPs.
\end{theorem}
Theorem~\ref{thm:main} follows by combining a number of known results. In particular, our main insight is to show that certain constructions of \emph{format-preserving} encryption yield quantum-secure PRPs.

\paragraph{Acknowledgments.}We would like to thank Scott Aaronson and Lijie Chen, both for needing a quantum-secure PRP in one of their results and for encouraging us to finally write this paper.  Without these motivations, the result would likely have remained unpublished for some time.

\subsection{Applications of Quantum PRPs.} \label{sec:apps}

To the best of our knowledge, the first proposed applications of quantum-secure PRPs are symmetric-key encryption secure in a strong quantum query model by Gagliardoni, H\"{u}lsing, and Schaffner~\cite{C:GagHulSch16}, and an efficient oracle separation for {\sf SZK} and {\sf BQP} by Aaronson and Chen~\cite{AC16}.

Subsequent to initially making our work public, quantum PRPs have been used in a number of important applications. Cryptographic applications include those of Zhandry~\cite{AC:Zhandry21} for analyzing the post-quantum security of the Sponge hash function, and of Bos, H\"{u}lsing, Renes, and van Vredendaal~\cite{TCHES:BHRv21} to study the post-quantum security of the XMSS signature scheme.

Quantum-secure PRPs have been used in a number of works on quantum pseudorandomness. For example, Behera, Brakerski, Sattath, and Shmueli~\cite{TCC:BBSS23} use quantum-secure PRPs to construct pseudorandomness with proofs of deletion and Lu, Qin, Song, Yao, and Zhao~\cite{LQSYZ24} use them to construct pseudorandom scramblers. Aaronson et al.~\cite{ITCS:ABFGVZ24}, Giurgica-Tiron and Bouland~\cite{GiuBou23}, and Arnon-Friedman, Brakerski, and Vidick~\cite{ArnBraVid23} use such PRPs to study quantum pseudo-entanglement. Chia and  Hung~\cite{ChiHun22} use quantum-secure PRPs to classically verify quantum depth. They were recently used by Ananth, Gulati, Kaleoglu, and Lin~\cite{EC:AGKL24} to construct pseudorandom isometries. Finally, in a very recent breakthrough work, quantum-secure PRPs were used by Huang and Ma~\cite{HuaMa24} to construct pseudorandom unitaries. Through our work, these results can all be based on quantum-immune one-way functions.

\subsection{Other Related Work}\label{sec:related}

\paragraph{Superposition security.} When considering classical cryptosystems in the quantum setting, one option is to keep the classical security experiment, but allow the adversary local quantum computation. A stronger security notion, however, is to update the security experiment itself to allow quantum queries. Such a stronger notion captures a wider class of attacks, and is often necessary if the protocol is to be used as part of a higher-level quantum protocol.

There have been numerous works considering such superposition security models. The previously mentioned work of~\cite{FOCS:Zhandry12} considers the case of quantum-secure pseudorandom \emph{functions}. Other examples include MPC~\cite{ICITS:DFNS13,EPRINT:ECKM20,EPRINT:LiuSahZha20}, message authentication/signatures~\cite{EC:BonZha13,C:BonZha13,C:KLLN16,C:GarYueZha17}, and encryption~\cite{C:BonZha13,C:BroJef15,C:GagHulSch16,EC:AMRS20}.

\paragraph{Quantum security of PRPs.} The quantum security of Feistel networks remains a largely open, very challenging problem. Kuwakado and Morii~\cite{KuwMor10} show that 3-round Feistel is insecure under forward-only quantum queries, in contrast to the classical setting where 3 rounds employing a suitable round function \emph{is} sufficient~\cite{LubRac88}. Hosoyamada and Iwata~\cite{AC:HosIwa19} claimed to prove that 4-round Feistel is quantum-secure under forward-only quantum queries, but it was very recently discovered that there was a bug in their proof~\cite{AC:CEJ24}. For non-adaptive forward-only queries,~\cite{AC:CEJ24} show that 4-round Feistel is still secure, but the adaptive query setting remains open. There are no positive results for Feistel networks in any number of rounds when allowing reverse queries. Our result remains the only provably secure PRP allowing inverse queries. 

Subsequent to the initial publication of our work,~\cite{Song17} sketched in a blog post how to construct quantum-secure PRPs using the rapid mixing of the Thorpe shuffle. Their main idea and construction are a special case of Theorem~\ref{thm:main}: namely, the rapid mixing of the Thorpe shuffle gives a format preserving encryption scheme sufficient for realizing quantum-secure PRPs.

\subsection{Our Techniques}

\paragraph{On adapting the classical approach.} The classical blueprint for constructing PRPs from one-way functions is usually described as follows.
\begin{itemize}
	\item H\r{a}stad, Impagliazzo, Levin, and Luby~\cite{HILL99} show how to use one-way functions to build pseudorandom generators (PRGs).  
	\item In turn, Goldreich, Goldwasser, and Micali~\cite{GolGolMic86} show how to use pseudorandom generators to build pseudorandom functions (PRFs), which are a relaxed notion of PRPs where the function is not required to be a permutation\footnote{There is also no efficient inverse function.}.  
	\item Finally, Luby and Rackoff~\cite{LubRac88} show that by plugging PRFs into a 3-round (respectively 4-round) Feistel Network, one obtains a PRP secure for forward-only queries (resp. forward and inverse queries).
\end{itemize}  

The obvious idea for constructing a quantum-secure PRP is to try to ``upgrade'' each step above to work quantumly:

\begin{itemize}
	\item {\bf Quantum immune one-way functions to quantum immune PRGs.}  Fortunately, the classical proof of security in~\cite{HILL99} can be lifted to the quantum setting. Intuitively, the classical security proof does not make any assumptions about the model of the adversary, and so works equally well when the adversary is classical or quantum. This result seems to be folklore, but is mentioned in~\cite{Aaronson2009,STOC:AarChr12,FOCS:Zhandry12}.
	\item {\bf Quantum immune PRGs to quantum-secure PRFs.}  Here, unfortunately, the classical model \emph{does} restrict the model of the adversary, as the adversary's queries are assumed to be classical in the classical proof.  The classical proof in~\cite{GolGolMic86} crucially uses the fact that the adversary only sees a polynomial number of \emph{classical} evaluations of the function. It is therefore no longer valid for proving security against quantum queries, who see superpositions over exponentially-many points.  Fortunately, Zhandry~\cite{FOCS:Zhandry12} gives a new and very different proof that does show the security of the GGM PRF against quantum algorithms making quantum queries.
	\item {\bf Quantum-secure PRFs to quantum-secure PRPs.}  Here again the classical model restricts the model of the adversary, so the classical proof in~\cite{LubRac88} is no longer valid. The classical security proof works as follows.  In the first step, the PRF in the Feistel network is replaced with a truly random function.  If we allow quantum queries to the PRP, we will need the PRF to be secure against quantum queries, but otherwise translating this step to the quantum setting is straightforward.  The next step, however, is problematic.  The next step is to show that the Feistel network, when instantiated with a truly random function, becomes indistinguishable from a truly random permutation.  Unfortunately, as with the GGM security proof above, the proof in the classical setting crucially relies on the fact that the adversary only ever gets to see a polynomial number of points, and completely breaks down if the adversary gets to ``see'' all exponentially many points.  As shown by Kuwakado and Morii~\cite{KuwMor10}, the 3-round Feistel network is insecure under quantum queries to only the forward directly, despite 3-rounds being sufficient for classical forward queries. It is entirely possible that with more rounds Feistel networks become indisitnguishable from random permutation under quantum queries, but no quantum security proof for Feistel networks is known\footnote{See Related Works below in Section~\ref{sec:related}.}, and any result along these lines would likely require a substantial reworking of Luby and Rackoff's analysis.
\end{itemize}

We will therefore devise a new approach to solving the third step above. Our construction is a simple combination of prior work. At a very high level, in any construction of a PRP from a PRF (quantum-secure or otherwise), the first step to prove security is to replace the PRF with a truly random function, invoking PRF security to show that change is indistinguishable. The second step is to take the resulting random object, and show that it ``looks like'' a truly random permutation. Importantly, this second step no longer utilizes the adversary's computational bounds, only the implied bounds on the number of queries. Essentially, what~\cite{LubRac88} show is that Feistel networks ``look like'' random permutations to algorithms making a bounded number of classical queries.

An analogous quantum statement for Feistel networks requires making query complexity arguments that seem beyond the reach of current techniques. Our key insight is that, in the second step, if the \emph{truth table} of the permutation (using a truly random function in place of the PRF) is statistically close to that of a random permutation, then this suffices for quantum security, even in the presence of quantum adversaries. Equivalently, if the construction (after replacing the PRF with a truly random function) is indistinguishable against algorithms that make \emph{classical} queries, but are allowed to query the \emph{entire domain}, then quantum security follows. Note that for this insight, it is crucial that we consider the information-theoretic object obtained by replacing the round-function PRF with a truly random function. Indeed, prior to replacing the PRF, the truth table \emph{cannot} look like a uniform permutation since there is not nearly enough entropy in the PRP key. 

``Full domain'' security is extremely strong, and it is well-known that the basic balanced Feistel networks such analyzed in~\cite{LubRac88} do not achieve such security. Indeed, a $k$-round Feistel network on $n$ bits consists of $k$ independent random ``round'' functions from $n/2$ bits to $n/2$ bits. After replacing the round functions with truly random functions, each round function has $(n/2)\times 2^{n/2}$ bits of entropy. The total entropy of the network is therefore at most $k\times (n/2)2^{(n/2)}$. On the other hand a random permutation will have roughly $n\times 2^n$ bits of entropy, far more than a $k$ round network can have for any polynomial $k$.

Fortunately, full domain PRPs have been widely studied in classical cryptography, under completely different motivations, namely in the context of \emph{format preserving encryption}~\cite{SAC:BRRS09}.  A core tool in format preserving encryption is a PRP where the domain can be very small, possibly even polynomial.  The challenge here is that, once the domain is polynomial, a polynomial-time adversary can query the function on the entire domain. For this reason, the balanced Feistel network cannot work in this regime, since they \emph{cannot} maintain security once queried on the entire domain. A fascinating line of work~\cite{STOC:Morris05,FSE:GraPor07,SAC:BRRS09,EPRINT:SteShi12,C:HoaMorRog12,C:RisYil13,EC:MorRog14} has shown how to build PRPs that (1) support small potentially polynomial-sized domains, and (2) support adversaries querying on the entire domain. By plugging in these results, we obtain a quantum-secure PRP from any quantum-secure PRF, and hence from any one-way function.

Now, in these works, since the domain is small, the overall construction can remain polynomial-time even if the run-time has a rather poor dependence on the domain size. Thus, small-domain full-domain PRPs are not necesarily enough to give large-domain quantum-secure  PRPs. However, in the interest of achieving practical constructions, the above sequence of works has shown that the dependence of the run-time on the domain size can be made essentially logarithmic. Thus, we can ``scale up'' these constructions  to the case of exponentially-large domains (that is, polynomial bit-length), and achieve a PRP secure against quantum queries even in the large-domain case, and the run-time remains polynomial.

\section{Preliminaries}

\label{sec:prelim}

A function $\epsilon=\epsilon(\lambda)$ is negligible if it is smaller than any inverse polynomial.  A \emph{classical} algorithm is said to be efficient if it runs in probabilistic polynomial time.  

\medskip

This work will make minimal use of quantum formalism.  Nonetheless, for completeness we recall the basics of quantum computation and quantum oracle queries.  The following is taken more or less verbatim from~\cite{C:BonZha13}.

\paragraph{Quantum Computation.} We give a short introduction to quantum computation.  A quantum system $A$ is a complex Hilbert space $\Hs$ together with and inner product $\langle\cdot|\cdot\rangle$.  The state of a quantum system is given by a vector $\qstate{}{}$ of unit norm ($\langle\psi|\psi\rangle=1$).  Given quantum systems $\Hs_1$ and $\Hs_2$, the joint quantum system is given by the tensor product $\Hs_1\otimes\Hs_2$.  Given $\qstate{1}{}\in\Hs_1$ and $\qstate{2}{}\in\Hs_2$, the product state is given by $\qstate{1}{}\qstate{2}{}\in\Hs_1\otimes\Hs_2$.  Given a quantum state $\qstate{}{}$ and an orthonormal basis $B=\{|b_0\rangle,...,|b_{d-1}\rangle\}$ for $\Hs$, a measurement of $\qstate{}{}$ in the basis $B$ results in the value $i$  with probability $|\langle b_i\qstate{}{}|^2$, and the quantum state collapses to the basis vector $|b_i\rangle$.  If $\qstate{}{}$ is actually a state in a joint system $\Hs\otimes\Hs'$, then $\qstate{}{}$ can be written as 

\[\qstate{}{}=\sum_{i=0}^{d-1}\;\alpha_i\; |b_i\rangle |\psi_i'\rangle\] 

for some complex values $\alpha_i$ and states $|\psi_i'\rangle$ over $\Hs'$.  Then, the measurement over $\Hs$ obtains the value~$i$ with probability $|\alpha_i|^2$ and in this case the resulting quantum state is $|b_i\rangle|\psi_i'\rangle$.

A unitary transformation over a $d$-dimensional Hilbert space $\Hs$ is a $d\times d$ matrix $\Um$ such that $\Um\Um^\dag=\Ident_d$, where $\Um^\dag$ represents the conjugate transpose.  A quantum algorithm operates on a product space $\Hs_{in}\otimes\Hs_{out}\otimes\Hs_{work}$ and consists of $n$ unitary transformations $\Um_1,...,\Um_n$ in this space.  $\Hs_{in}$ represents the input to the algorithm, $\Hs_{out}$ the output, and $\Hs_{work}$ the work space.  A classical input $x$ to the quantum algorithm is converted to the quantum state $|x,0,0\rangle$.  Then, the unitary transformations are applied one-by-one, resulting in the final state 

\[\qstate{x}{} = \Um_n...\Um_1 |x,0,0\rangle\enspace .\]

The final state is then measured, obtaining the tuple $(a,b,c)$ with probability $\left|\langle a,b,c\qstate{x}{}\right|^2$.  The output of the algorithm is $b$.  We say that a quantum algorithm is efficient if each of the unitary matrices $\Um_i$ come from some fixed basis set, and $n$, the number of unitary matrices, is polynomial in the size of the input.

\paragraph{Quantum-accessible Oracles.}  We will implement an oracle $O:\Xs\rightarrow\Ys$ by a unitary transformation $\Om$ where 

\[\Om |x,y,z\rangle = |x,y+O(x),z\rangle\]

where $+:\Xs\times\Xs\rightarrow\Xs$ is some group operation on $\Xs$.  Suppose we have a quantum algorithm that makes quantum queries to oracles $O_1,...,O_q$.  Let $|\psi_0\rangle$ be the input state of the algorithm, and let $\Um_0,...,\Um_q$ be the unitary transformations applied between queries.  Note that the transformations $\Um_i$ are themselves possibly the products of many simpler unitary transformations.  The final state of the algorithm will be 

\[\Um_q \Om_q ... \Um_1 \Om_1 \Um_0 |\psi_0\rangle\enspace .\]

We can also have an algorithm make classical queries to $O_i$.  In this case, the input to the oracle is measured before applying the transformation $\Om_i$.  We call a quantum oracle algorithm efficient if the number of queries $q$ is a polyomial, and each of the transformations $\Um_i$ between queries can be written as the product polynomially many unitary transformations from some fixed basis set.

\paragraph{Conventions for this paper.}  For this paper, all cryptographic primitives discussed will be implemented by efficient \emph{classical} algorithms.  We will use $\CComp$ and $\QComp$ to distinguish between classical and quantum adversaries, and we will use $\CQuery$ and $\QQuery$ to distinguish between adversaries making classical or quantum queries.

\subsection{Cryptographic Primitives}

All algorithms and adversaries will take as input a security parameter $\lambda$.  We will use the convention that $\lambda$ is specified in binary, and algorithms are efficient if they run in time polynomial in $\lambda$ (which is exponential in the bit-length of $\lambda$).

\begin{definition}A $\CComp$- (resp. $\QComp$-) \emph{one-way} function is an efficient classical function $\OWF:\{0,1\}^\lambda\rightarrow\{0,1\}^*$ such that, for any efficient classical (resp. quantum) adversary $\adv$, the probability that $\adv$ inverts $\OWF$ on a random input is negligible.  That is, there exists a negligible $\negl(\lambda)$ such that \[\Pr[\OWF(\;\adv(\lambda,\OWF(x))\;)=\OWF(x):x\gets\{0,1\}^\lambda]<\negl(\lambda)\enspace .\]
\end{definition}

\begin{definition}For a pair $(\Comp,\Query)\in\{(\CComp,\CQuery),\allowbreak(\QComp,\CQuery),\allowbreak(\QComp,\QQuery)\}$,\footnote{Note that it does not make sense to consider classical adversary's that make quantum queries.} A $(\Comp,\Query)$-\emph{pseudorandom function} (PRF),  is a family of efficient classical functions $\PRF_{m,n}:\{0,1\}^\lambda\times\{0,1\}^m\rightarrow\{0,1\}^n$ such that the following holds.  For any polynomially bounded $m=m(\lambda)$ and $n=n(\lambda)$, and any efficient adversary $\adv$, $\adv$ cannot distinguish $\PRF_{m,n}(k,\cdot)$ for a random $k\gets\{0,1\}^\lambda$ from a truly random function $O:\{0,1\}^m\rightarrow\{0,1\}^n$.  That is, there exists a negligible $\negl(\lambda)$ such that \[\left|\Pr[\adv^{\PRF_{m,n}(k,\cdot)}(\lambda)=1:k\gets\{0,1\}^\lambda]-\Pr[\adv^{O(\cdot)}(\lambda)=1]\right|<\negl(\lambda)\enspace .\]

Here, $O$ is chosen at random from the set of all functions from $\{0,1\}^m$ into $\{0,1\}^n$.  If $(\Comp,\Query)=(\CComp,\CQuery)$, $\adv$ is restricted to being an efficient classical algorithm making classical queries to $O$.  If $\Comp=\QComp$, then $\adv$ is allowed to be a quantum algorithm.  In this case, if $\Query=\CQuery$, $\adv$, despite being quantum, is still required to make classical queries.  If $\Query=\QQuery$, $\adv$ is allowed to make quantum queries to $O$.  

We will often abuse notation when $m$ and $n$ are clear from context and write $\PRF$ to denote $\PRF_{m,n}$.
\end{definition}

We note that PRFs domains and co-domains are \emph{monotone}, in the sense that a PRF on for large domain and range gives a PRF for small domain and range, simply by hard-coding some bits of the inputs and discarding some bits of the output.  Therefore, if a PRF is secure for large domain/range sizes (say, polynomial $m$ and $n$), it can also easily be made secure for small domain and range sizes (say, logarithmic or polylogarithmic $m$ and $n$).  

We recall the following theorem, combining the results~\cite{GolGolMic86,HILL99,FOCS:Zhandry12}.  

\begin{theorem}[Combination of~\cite{GolGolMic86,HILL99,FOCS:Zhandry12}]\label{thm:PRF} If $\CComp$-one-way functions exist, then so do $(\CComp,\CQuery)$-PRFs.  Moreover, if $\QComp$-one-way functions exist, then so do $(\QComp,\QQuery)$-PRFs.
\end{theorem}
		
The classical version follows from~\cite{GolGolMic86,HILL99}.  The reductions in those works are actually independent of the computational model, so the proofs also show that $\QComp$-one-way functions imply $(\QComp,\CQuery)$-PRFs.  The analysis does not extend to the $\QQuery$ setting.  Instead, the work of~\cite{FOCS:Zhandry12} shows how to modify the proof to get the quantum part of Theorem~\ref{thm:PRF}.





\section{Pseudorandom Permutations}

\label{sec:fpc}

\begin{definition}For a pair $(\Comp,\Query)\in\{(\CComp,\CQuery),\allowbreak(\QComp,\CQuery),\allowbreak(\QComp,\QQuery)\}$ and $\Domain\in\{\LargeD,\AnyD\}$, a $(\Comp,\Query,\Domain)$-\emph{pseudorandom permutation} (PRP),  is a family of efficient classical function pairs $\PRP_{o}:\{0,1\}^\lambda\times\{0,1\}^o\rightarrow\{0,1\}^o$ and $\PRP_{o}^{-1}:\{0,1\}^\lambda\times\{0,1\}^o\rightarrow\{0,1\}^o$ such that the following holds.  
\begin{itemize}
	\item First, for every key $k$ and integer $o$, the functions $\PRP_o(k,\cdot)$ and $\PRP_o^{-1}(k,\cdot)$ are inverses of each other.  That is, $\PRP_o^{-1}(k,\PRP_o(k,x))=x$ for all $o,k,x$.  This implies that $\PRP_o(k,\cdot)$ is a permutation.
	\item Second, for any polynomially-bounded $o=o(\lambda)$ and any efficient adversary $\adv$, $\adv$ cannot distinguish $\PRP_{o}(k,\cdot)$ for a random $k\gets\{0,1\}^\lambda$ from a truly random \emph{permutation} $P:\{0,1\}^o\rightarrow\{0,1\}^o$.  We consider the strong variant where $\adv$ has access to both $P$ and $P^{-1}$.  That is, there exists a negligible $\negl(\lambda)$ such that \[\left|\Pr[\adv^{\PRP_{o}(k,\cdot),\PRP_{o}^{-1}(k,\cdot)}(\lambda)=1:k\gets\{0,1\}^\lambda]-\Pr[\adv^{P(\cdot),P^{-1}(\cdot)}(\lambda)=1]\right|<\negl(\lambda)\enspace.\]
	
	Here, $P$ is chosen at random from the set of all permutations on $\{0,1\}^o$.  If $(\Comp,\Query)=(\CComp,\CQuery)$, $\adv$ is restricted to being an efficient classical algorithm making classical queries to $P,P^{-1}$.  If $\Comp=\QComp$, then $\adv$ is allowed to be a quantum algorithm.  In this case, if $\Query=\CQuery$, $\adv$, despite being quantum, is still required to make classical queries.  If $\Query=\QQuery$, $\adv$ is allowed to make quantum queries to $P,P^{-1}$.  
	\item Finally, if $\Domain=\LargeD$, we only require security to hold for $o$ that are upper- and lower-bounded by a polynomial.  If $\Domain=\AnyD$, we allow for arbitrary polynomially-upper-bounded $o$ (so $o$ could be, for example, logarithmic in this case).  Note that a PRP for domain $\{0,1\}^o$ does not give a PRP for domain $\{0,1\}^{o'}$ for $o'<o$ by fixing bits, since $\PRP_o$ is not guaranteed to be a permutation on $\{0,1\}^{o'}$.  Therefore, we make a distinction between PRPs that only work for sufficiently large domains ($\LargeD$), and those that remain secure for both large and small domains $(\AnyD)$.
\end{itemize}	
We will often abuse notation when $o$ is clear from context and write $\PRP$ to denote $\PRP_{o}$.
\end{definition}

\subsection{Function to Permutation Converters}

In this section, we define a information theoretic object called a \emph{function to permutation converter} (FPC), which roughly takes a random function and transforms it into a random permutation.  Such FPCs are implicitly used in many constructions of PRPs, and basically every construction of a PRP from general one-way functions.

\begin{definition} Let $\Query\in\{\CQuery,\QQuery\}$, $\Domain\in\{\LargeD,\AnyD\}$.  Fix a family $\Qs$ of functions over $o,\lambda$.  An $(\Query,\Domain,\Qs)$-FPC is a sequence of pairs of efficient classical oracle algorithms $\FPC_o,\FPCInv_o$ where:

\begin{itemize}
	\item $\FPC_o,\FPCInv_o$ take as input $\lambda$  and string $x\in\{0,1\}^o$, and output a string $y\in\{0,1\}^o$
	\item There exist polynomials $m(o,\lambda),n(o,\lambda)$ such that $\FPC_o,\FPCInv_o$ make queries to a function $O:\{0,1\}^m\rightarrow\{0,1\}^n$. 
	\item $\FPC_o,\FPCInv_o$ are efficient, running in time $\poly(o,\lambda)$. In particular, they make at most $\poly(o,\lambda)$ queries to $O$.
	\item $\FPC_o,\FPCInv_o$ are inverses: $\FPCInv_o^O(\lambda,\FPC_o^O(\lambda,x))=x$ for all $x\in\{0,1\}^o$ and all oracles $O$.
	\item $\FPC_o,\FPCInv_o$ are indistinguishable from a random permutation and its inverse, given query access.  That is, let $o(\lambda)$ be any polynomially bounded function.  If $\Domain=\LargeD$, we will restrict to $o$ being lower-bounded by a polynomial as well. Let $q(\lambda)=q(o(\lambda),\lambda)$ be any function in $\Qs$.  Let $\adv$ be any (possibly inefficient) adversary that makes at most $q$ queries to its oracles, where if $\Query=\QQuery$ the queries are allowed to quantum, and if $\Query=\CQuery$, the queries are restricted to being classical.  Then there exists a negligible $\negl(\lambda)$ such that:
	\[\left|\Pr[\adv^{\FPC_o^O(\lambda,\cdot),\FPCInv_o^O(\lambda,\cdot)}(\lambda)=1]-\Pr[\adv^{P,P^{-1}}(\lambda)=1]\right|<\negl(\lambda)\enspace.\]
\end{itemize}
\end{definition}

\subsection{The Main Lemma}
Here we prove that FPCs are sufficient to build PRPs.

\begin{lemma}\label{lem:main} Let $\Qs$ be the set of all polynomials.  If $(\Comp,\Query)$-PRFs exist and $(\Query,\Domain,\Qs)$-FPCs exist, then $(\Comp,\Query,\Domain)$-PRPs exist.
\end{lemma}

We note that the $(\Comp,\Query,\Domain)=(\CComp,\CQuery,\Domain)$ version of this lemma is implicit in essentially all constructions of classically secure PRPs from general PRFs.  The proof is a straightforward adaptation to handle quantum queries.

\begin{proof}We prove the $(\QComp,\QQuery,\AnyD)$ version, the other versions being similar.  The construction is simple: $\PRP_o(k,x)=\FPC_o^{\PRF_{m,n}(k,\cdot)}(\lambda,x)$ and $\PRP^{-1}_o(k,x)=\FPCInv_o^{\PRF_{m,n}(k,\cdot)}(\lambda,x)$ where $\lambda=|k|$.
	
Security is proved by a sequence of hybrids.

\paragraph{Hybrid 0.}  This is the case where the adversary is given the oracles $P(x)=\FPC_o^{\PRF_{m,n}(k,\cdot)}(\lambda,x)$ and $P^{-1}(x)=\FPCInv_o^{\PRF_{m,n}(k,\cdot)}(\lambda,x)$.  

\paragraph{Hybrid 1.} In this case, we switch $\PRF$ to be random, so that the adversary is given the oracles $P(x)=\FPC_o^{O(\cdot)}(\lambda,x)$ and $P^{-1}(x)=\FPCInv_o^{O(\cdot)}(\lambda,x)$ for a random function $O:\{0,1\}^m\rightarrow\{0,1\}^n$.  

Suppose $\adv$ distinguishes {\bf Hybrid 0} from {\bf Hybrid 1} with non-negligible probability.  Then we can construct a PRF adversary $\advB^O(\lambda)=\adv^{\FPC_o^{O}(\lambda,\cdot),\FPCInv_o^{O}(\lambda,\cdot)}(\lambda)$.  If $O(\cdot)=\PRF_{m,n}(k,\cdot)$ for a random $k$, then the view of $\adv$ is identical to {\bf Hybrid 0}.  Likewise, if $O(\cdot)$ is a truly random function, then the view of $\adv$ is identical to {\bf Hybrid 1}.  Moreover, $\advB$ can answer any quantum query made by $\adv$ by making a polynomial number of quantum queries to its own oracle.  Therefore, $\advB$ is an efficient quantum algorithm.  Moreover, its advantage in breaking the security of $\PRF$ is identical to $\adv$'s distinguishing advantage between {\bf Hybrid 0} and {\bf Hybrid 1}.  This contradicts our assumption that $\PRF$ is $(\QComp,\QQuery)$ secure.  

Therefore, {\bf Hybrid 0} and {\bf Hybrid 1} are indistinguishable, except with negligible probability.

\paragraph{Hybrid 2.}  In this hybrid, the adversary is given truly random permutation oracles $P,P^{-1}$.  Indistinguishability between {\bf Hybrid 1} and {\bf Hybrid 2} follows immediately from the $(\QQuery,\AnyD,\Qs)$ security of $\FPC,\FPCInv$.  

Thus, {\bf Hybrid 0} and {\bf Hybrid 2} are indistinguishable, demonstrating the security of $\PRP$.\end{proof}

\subsection{Constructions of FPCs}

From Theorem~\ref{thm:PRF}, we already have PRFs from one-way functions.  It remains to demonstrate a suitable FPC.  Luby and Rackoff's~\cite{LubRac88} proof essentially shows the following:
\begin{lemma}[Implicit in~\cite{LubRac88}]\label{lem:LR} Let $\Qs$ be the set of polynomial-bounded functions.  Then $(\CQuery,\allowbreak\LargeD,\allowbreak\Qs)$-FPCs exist.
\end{lemma}

\noindent This gives us the following corollary, showing that (large-domain) PRPs secure against classical queries exist:
\begin{corollary}\label{cor:largeD} For $\Comp\in\{\CComp,\QComp\}$, if $\Comp$-one-way functions exist, then $(\Comp,\allowbreak\CQuery,\allowbreak\LargeD)$-PRPs exist.
\end{corollary}
\begin{proof}Assuming $\Comp$-one-way functions exist, by Theorem~\ref{thm:PRF}, so do $(\Comp,\CQuery)$-PRFs. Lemma~\ref{lem:LR} shows the existence of $(\CQuery,\LargeD,\Qs)$-FPCs, where $\Qs$ is the set of polynomially-bounded functions. Plugging the PRF and FPC into Lemma~\ref{lem:main} with $\Query=\CQuery$ and $\Domain=\LargeD$ then gives the desired $(\Comp,\CQuery,\LargeD)$-PRPs.
\end{proof}

Luby Rackoff's construction uses Feistel networks.  If we could show that Feistel networks are also $(\QQuery,\LargeD,\Qs)$-FPCs, then we would immediately get $(\QComp,\QQuery,\LargeD)$-PRPs from any $\QComp$-one-way function as desired.  Unfortunately, we do not know how to show this. It is known that more rounds are necessary than in the classical setting~\cite{KuwMor10}, but it is unknown whether any number of rounds are sufficient quantumly. We therefore leave the quantum security of the Feistel network as an important open question.

Instead, we observe that if $q=2^o$, then even a classical algorithm can query the entire domain of $\FPC_o$, obtaining the entire truth table for the function (as well as its inverse $\FPCInv_o$).  Then it can simulate any query, quantum or otherwise.  Thus, in this regime, there is no distinction between classical and quantum FPCs.  Note that this regime still makes sense even though an adversary making $2^o$ queries is inefficient, since FPC security is defined for computationally unbounded adversaries.  This motivates the following definition:

\begin{definition} The family $(\FPC_o,\FPCInv_o)$ is a $\Domain$-Full Domain FPC if there exists a set $\Qs$ containing a function $q(o,\lambda)$ such that $q(o,\lambda)\geq 2^o$, such that $(\FPC_o,\FPCInv_o)$ is a $(\CQuery,\Domain,\Qs)$-FPC.
\end{definition}

\begin{lemma}\label{lem:fulldomain2quant}  Any $\Domain$-Full Domain FPC is also a $(\QQuery,\Domain,\Qs)$-FPC, for \emph{any} class of functions $\Qs$.
\end{lemma}
\begin{proof}Let $\adv$ be any (possibly inefficient) algorithm making quantum queries to a $\Domain$-Full Domain FPC, and consider the quantity
	\[\epsilon(\lambda):=\left|\Pr[\adv^{\FPC_o^O(\lambda,\cdot),\FPCInv_o^O(\lambda,\cdot)}(\lambda)=1]-\Pr[\adv^{P,P^{-1}}(\lambda)=1]\right|\enspace .\]
Our goal is to show that $\epsilon$ must be negligible, regardless of how many (quantum) queries $\adv$ makes. Now let $\advB$ be the following inefficient classical query algorithm: using $2^o$ queries, it queries the entire domain of its oracle. $\advB$ then uses the query responses to build the truth table $T$ for its oracle. It then runs $\adv^{T,T^{-1}}$, where the oracle $T$ is a quantum-accessible classical oracle defined by the truth table $T$, and $T^{-1}$ is its (quantumly-accessible) inverse. $\advB$ outputs whatever $\adv$ outputs. The oracle $T$ seen by $\adv$ is exactly the same as the oracle seen by $\advB$. Hence, we have that 
\[\left|\Pr[\advB^{\FPC_o^O(\lambda,\cdot),\FPCInv_o^O(\lambda,\cdot)}(\lambda)=1]-\Pr[\advB^{P,P^{-1}}(\lambda)=1]\right|=\epsilon(\lambda)\enspace .\]
Since $\Domain$-Full Domain security applies to algorithms making $2^o$ queries, and hence applies to $\advB$, we have that $\epsilon$ is negligible, as desired.
\end{proof}

\medskip

Next, we observe that the card shuffles at the heart of constructions of format preserving encryption give us Full-Domain FPCs.  The following is adapted from~\cite{STOC:Morris05,FSE:GraPor07,EPRINT:SteShi12,C:HoaMorRog12,C:RisYil13,EC:MorRog14}.

\begin{lemma}[Implicit in~\cite{STOC:Morris05,FSE:GraPor07,EPRINT:SteShi12,C:HoaMorRog12,C:RisYil13,EC:MorRog14}]\label{lem:fulldomain} $\AnyD$-Full Domain FPCs exist.
\end{lemma}
The main difference between these various works is how efficient the constructions are. The goal of these works is to use Lemma~\ref{lem:fulldomain} to give the following improvement to Corollary~\ref{cor:largeD}, namely achieving security for arbitrary domains, which we adapt here:
\begin{corollary}For $\Comp\in\{\CComp,\QComp\}$, if $\Comp$-one-way functions exist, then $(\Comp,\allowbreak\CQuery,\allowbreak\AnyD)$-PRPs exist.
\end{corollary}
\begin{proof}The proof is identical to Corollary~\ref{cor:largeD}, but with with $\LargeD$ replaced with $\AnyD$ and Lemma~\ref{lem:LR} replaced with Lemma~\ref{lem:fulldomain}.
\end{proof}

\subsection{Putting it All Together}

Combining Theorem~\ref{thm:PRF} with Lemmas~\ref{lem:main}, \ref{lem:fulldomain2quant}, and~\ref{lem:fulldomain}, quantum-secure PRPs exist, giving a slightly more precise version of Theorem~\ref{thm:main}:

\begin{theorem} If $\QComp$-one-way functions exist, then so do $(\QComp,\QQuery,\AnyD)$-PRPs.
\end{theorem}
\begin{proof}Assuming $\QComp$-one-way functions exist, by Theorem~\ref{thm:PRF}, so do $(\QComp,\QQuery)$-PRFs. Lemma~\ref{lem:fulldomain} shows the existence of $\AnyD$-Full Domain FPCs, which by Lemma~\ref{lem:fulldomain2quant} are also $(\QQuery,\Domain,\Qs)$-FPCs for any class of functions $\Qs$; therefore set $\Qs$ to be the set of all polynomials. Plugging the PRF and FPC into Lemma~\ref{lem:main} with $\Query=\QQuery$ and $\Domain=\AnyD$ then gives the desired $(\QComp,\QQuery,\AnyD)$-PRPs.
\end{proof}

\bibliographystyle{alpha}
\bibliography{extracted}

\end{document}